\documentclass[letter]{aa}

\usepackage{graphicx}
\usepackage{txfonts}
\usepackage{hyperref}
\usepackage{graphicx}
\usepackage{gensymb}
\usepackage{multirow}
\usepackage{amsmath}
\usepackage{tablefootnote}
\usepackage{ulem}
\begin{document}

   \title{Evidence for globular cluster collapse after a dwarf-dwarf merger: A potential nuclear star cluster in formation}

   \titlerunning{The path to the formation of a nuclear star cluster?}
   {}
   \authorrunning{Román et al.}

   \author{
          J.~Román \inst{1,2,3} \thanks{\email{jromanastro@gmail.com}}
          \and 
          P.~M.~Sánchez-Alarcón \inst{2,3}
          \and
          J.~H.~Knapen \inst{2,3}
          \and 
          R. Peletier \inst{1}
          }

   \institute{Kapteyn Astronomical Institute, University of Groningen, PO Box 800, 9700 AV Groningen, The Netherlands
   \and Instituto de Astrof\'{\i}sica de Canarias, c/ V\'{\i}a L\'actea s/n, E-38205, La Laguna, Tenerife, Spain
   \and Departamento de Astrof\'{\i}sica, Universidad de La Laguna, E-38206, La Laguna, Tenerife, Spain
   }
   \date{\today}

% \abstract{}{}{}{}{} 
% 5 {} token are mandatory
 
  \abstract
   {Direct observational evidence for the creation of nuclear star clusters (NSCs) is needed to support  the proposed scenarios for their formation. We analysed the dwarf galaxy UGC~7346, located in the peripheral regions of the Virgo Cluster, to highlight a series of properties that indicate the formation of a NSC caught in its earlier stages. First, we report on remnants of a past interaction in the form of diffuse streams or shells, suggesting a recent merging of two dwarf galaxies with a 1:5 stellar mass ratio. Second, we identify a number of globular cluster (GC) candidates that are broadly compatible in colour with the main component that is both more extended and more massive. Strikingly, we find these GCs candidates to be highly concentrated towards the centre of the galaxy (R$_{GC}$~=~0.41~R$_{e}$). We {suggest} that the central concentration of the GCs is likely produced by the dynamical friction of this merger. This would make UGC~7346 a unique case of a galaxy caught in the earlier stages of NSC formation. The formation of NSCs due to collapse of GCs by dynamical friction in dwarf mergers would provide a natural explanation of the environmental correlations found for the nucleation fraction for early-type dwarf galaxies, whereby denser environments host galaxies with a higher nucleation fraction.}
   
   %If future work confirms dwarf-dwarf merger as a viable channel of NSC formation, the expected high frequency of dwarf-dwarf mergers in $\Lambda$-CDM would provide a natural explanation to the observed environmental correlations found for the nucleation fraction for early-type dwarf galaxies, in which denser environments host galaxies with a higher nucleation fraction.}

%The formation of NSCs induced by mergers, and subsequent collapse of the GCs by dynamical friction, would provide a natural explanation to the observed environmental correlations found for the nucleation fraction for early-type dwarf galaxies, in which denser environments host galaxies with a higher nucleation fraction.


   \maketitle
%
%-------------------------------------------------------------------

\section{Introduction}

Nuclear star clusters (NSCs) are extremely compact stellar structures located in the central regions of so-called nucleated galaxies \citep[for a recent review, see][]{2020A&ARv..28....4N}. The presence of NSCs in galaxies was first noticed in early studies in the Virgo cluster \citep{1983ApJS...53..375R, 1983AJ.....88..804C, 1987AJ.....94..251B}. A large number of galaxies in different environments have since been observed to understand the properties of NSCs. The fraction of nucleated galaxies is strongly dependent on their mass and morphology, reaching a maximum nucleation fraction of about 80\% for $M_\star$~=~10$^{9} M_\odot$, decreasing for higher and lower masses \citep[][and references therein]{2020A&ARv..28....4N}. The fraction of nucleated galaxies also depends on the environment, with the fraction higher in denser environments \citep[][]{2019ApJ...878...18S, 2022A&A...664A.167S, 2022ApJ...927...44C}. The intrinsic properties of NSCs correlate strongly with those of their host galaxy \citep{2003ApJ...582L..79B, 2006ApJS..165...57C, 2006ApJ...644L..21F, 2007ApJ...665.1084B}. 

Despite exhaustive studies trying to unravel how NSCs are formed, a complete picture of their formation mechanism has not yet emerged \citep[][]{2020A&ARv..28....4N}, however, two different scenarios have been deemed as plausible. The coalescence of globular clusters (GCs) by dynamical friction is the most widely established  mechanism for the formation of NSCs \citep[e.g.,][]{1975ApJ...196..407T, 1993ApJ...415..616C, 2001ApJ...552..572L, 2000ApJ...543..620O}. However, the presence of young stellar populations in NSCs, especially in late-type galaxies \citep[e.g.,][]{2005ApJ...631..280B, 2013ApJ...764..155L, 2014MNRAS.441.3570G, 2015AJ....149..170C} but also in early-type galaxies \citep[][]{2017ApJ...836..237N, 2019ApJ...872..104N}, suggests a potential contribution on the part of in situ star formation in NSCs. There is a lot of evidence that indicates a transition in the properties of NSCs in galaxies for a mass threshold of $M_\star$~=~10$^{9} M_\odot$ \citep[see][and references therein]{2020A&ARv..28....4N}. Galaxies below this mass would have a NSC formation mechanism dominated by GC coalescence, while for masses above this threshold, in situ formation would be dominant. {This statement is supported by a recent analysis of the star formation histories of NSCs by \cite{2021A&A...650A.137F}.}  

Despite the clear evidence suggesting these formation channels for NSCs, there is still no direct observational evidence showing a case where a NSC is caught being formed \citep[however, see][for the case of a nuclear merger in its later stages]{2021MNRAS.503..594S}. The very short timescales on which NSCs are expected to be formed \citep[e.g.][]{2008ApJ...681.1136C, 2008MNRAS.388L..69C} lower the chances of observing a merger of GCs into a NSC. Additionally, some properties of NSCs remain to be explained under these two formation mechanisms. For instance, the correlation between the fraction of nucleated galaxies and the environment \citep[e.g. ][]{2019ApJ...878...18S, 2022A&A...664A.167S, 2022ApJ...927...44C} offers no easy explanation under these proposed mechanisms, as they are internal to the hosts. 

In this letter, we base our analysis on the {dwarf elliptical (dE)} galaxy UGC~7346. We argue that it is a case of a dwarf-dwarf merger inducing the collapse of its GCs and, thus, we consider it to be in the early stages of NSC formation. We assume a distance for UGC~7346 of 16.5 Mpc \citep{2007ApJ...655..144M}, implying a distance modulus of $m$~-~$M$~=~31.09~mag and a scale of 0.080~kpc~arcsec$^{-1}$. All photometric quantities are extinction-corrected following \cite{2011ApJ...737..103S}. We use the AB photometric system throughout this work.

\section{Data}

We use publicly available data from the Dark Energy Camera Legacy Survey \citep[DECaLS;][]{2019AJ....157..168D}. This survey provides deep \textit{g}, \textit{r,} and \textit{z} band images. We downloaded stamps centred on UGC~7346 in all bands via webpage of the survey\footnote{\url{https://www.legacysurvey.org}}. We use the seventh data release. The limiting surface brightness calculated following the method by \cite{2020A&A...644A..42R} are 28.9, 28.2, and 27.3 mag arcsec$^{-2}$ [3$\sigma$; 10$\times$10 arcsec] for the \textit{g}, \textit{r,} and \textit{z} bands, respectively. 

%The 5$\sigma$ point source detection as provided by the survey is 24.7, 23.6 and 22.8 mag for the \textit{g}, \textit{r} and \textit{z} bands respectively.

We explored the existence of other datasets, but found no data available in the CFHT Science Archive or the Hubble Legacy Archive. We found data from Pan-STARRS, however because they are shallower and less optimally processed compared to DECaLS data, we discard their use.

\section{Analysis}

\subsection{Photometric decomposition}\label{sec:Models}

\begin{figure*}
\centering
        \includegraphics[width=1.0\textwidth]{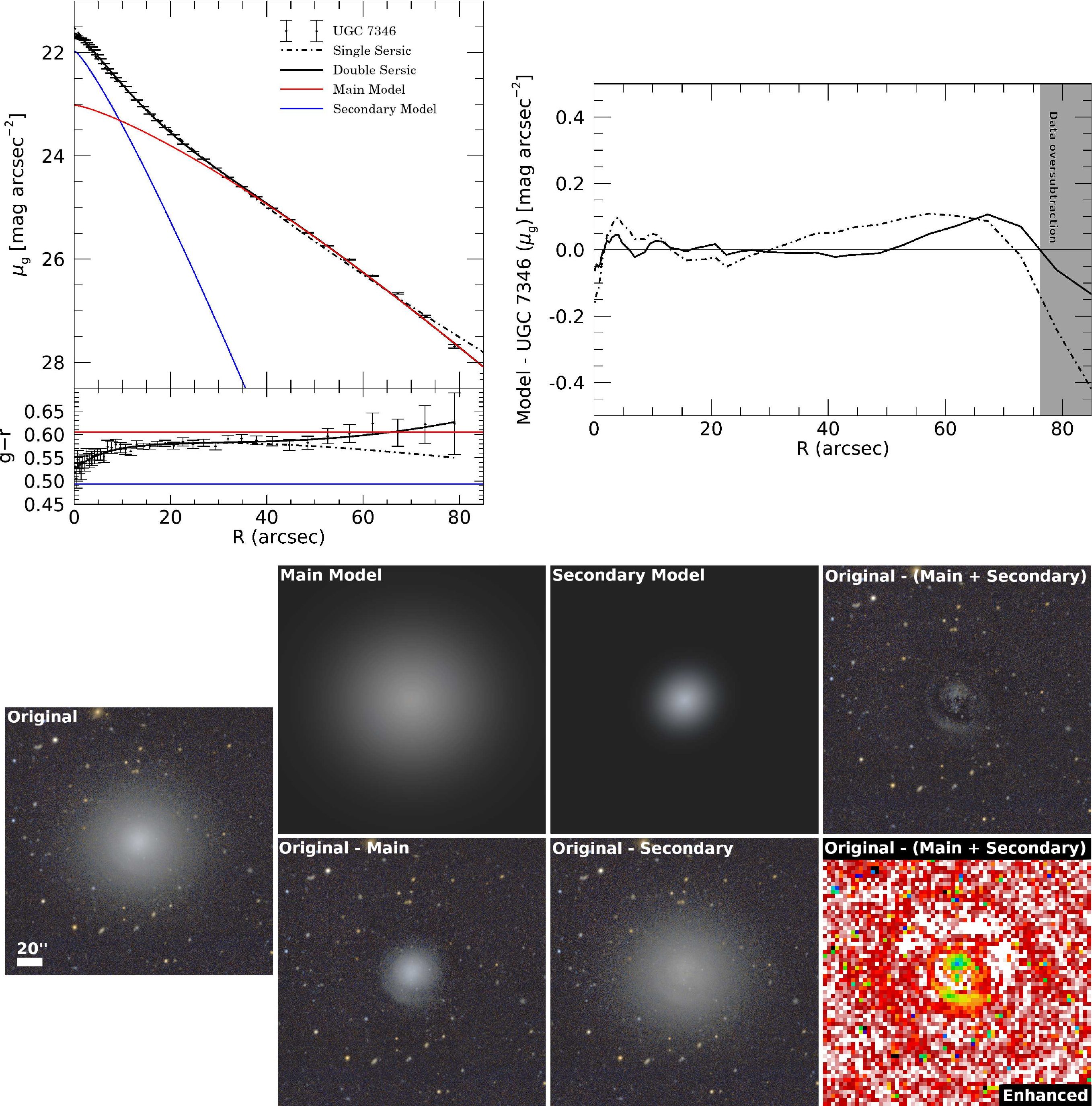}
    \caption{Photometric analysis and decomposition of UGC~7346. Upper left panel: Surface brightness profile in \textit{g} band and \textit{g}-\textit{r} colour profile for UGC~7346 and models (see text). Upper right panel: Residual after subtracting the surface brightness profile of UGC~7346 from the models. Bottom: Modelling of UGC~7346 in two different Sersic components. The left image is the original image. The upper centre images are the two fitted Sersic models, showing the corresponding residuals of these models in the lower central images. The result of subtracting the sum of the two models is shown in the right images. The enhanced residual image is produced by binning the \textit{g}+\textit{r} residual image to a size of 10x10 original pixels and using a high contrast false colour, allowing a greater appreciation of the lower surface brightness structures. The rest of the colour images are constructed using \textit{g} and \textit{r} bands. All images have a size of 3.6~$\times$~3.6 arcmin. North is up, east to the left.}
    \label{fig:Panel_photo}
\end{figure*}

The surface brightness profile of UGC~7346 is characterised by a central concentration of bluer light and a redder flat profile in the outer regions (Fig. \ref{fig:Panel_photo}). We modelled the structure of UGC~7346 with a structural analysis with Sersic \citep{1968adga.book.....S} models using the IMFIT software \citep{2015ApJ...799..226E}. Two different settings were employed: single Sersic and double Sersic models. {For the fitting, we first calculated the ellipticity, position angle, and central coordinates of the models on the \textit{g}+\textit{r}+\textit{z} image. These parameters are subsequently fixed for the fitting of the individual bands, leaving the rest of the parameters free. The fitting was carried out with a chi-square minimization using the IMFIT default parameters.}

In Fig. \ref{fig:Panel_photo}, we plot the \textit{g} band surface brightness profile of UGC~7346, together with the fitted single and double Sersic models, as well as the \textit{g}-\textit{r} colour profile (we omitted the \textit{r}-\textit{z} colour due to the low signal to noise of the \textit{z}-band). The single Sersic model produces slight deviations from the profile of UGC~7346, while the double Sersic model fits it better. In order to appreciate more clearly the differences in the surface brightness profiles between the fitted models and UGC~7346, we subtracted the models from the profile of UGC~7346 (Fig. \ref{fig:Panel_photo}, right panel). The single Sersic model has larger deviations from the UGC~7346 profile with a v-shape feature with vertex at approximately 20 arcsec. On the other hand, the difference with the double Sersic model remains approximately constant around 0, with slight undulations, which as we will discuss later are due to the substructure in UGC~7346 has underneath. The models deviate significantly at high radius (> 80 arcsec), probably due to sky oversubtraction in the Legacy Survey data.

Regarding the \textit{g}-\textit{r} colour profile, both the single and double Sersic models reproduce the central colour of UGC~7346. However, in the outer parts, the single model fails to reproduce the redder colour of UGC~7346 beyond 50 arcsec. Based strictly on the analysis of the surface brightness profiles, the double Sersic model fits more accurately the structure of UGC~7346. As we  discuss in the following, there are supplementary arguments to support a double Sersic component in UGC~7346.

\begin{table}
\begin{center}
\begin{tabular}{ccc}
Parameter & Main model & Secondary model \\ 
\hline
R$_{e}$ [arcsec] & 31.9 $\pm$ 1.7 & 11.0 $\pm$ 1.2\\
S\'ersic index & 0.78 $\pm$ 0.06 & 0.84 $\pm$ 0.05\\
Axis ratio & 0.94 $\pm$ 0.01 & 0.91 $\pm$ 0.01 \\
Position angle [deg] & 89 $\pm$ 3 & 125 $\pm$ 7 \\
\textit{g}-band [mag] & 14.33 $\pm$ 0.04 & 15.75 $\pm$ 0.08\\
\textit{r}-band [mag] & 13.73 $\pm$ 0.04 & 15.25 $\pm$ 0.08\\
\textit{z}-band [mag] & 13.46 $\pm$ 0.11 & 14.61 $\pm$ 0.20\\
L$_{g}$ [mag] & -16.76 $\pm$ 0.04 & -15.34 $\pm$ 0.08\\
M$_\star$ [M$_\odot$] & 1.2 $\pm$ 0.2 $\times$ 10$^{9}$ & 2.4 $\pm$ 0.8 $\times$ 10$^{8}$\\
\hline
\newline
\end{tabular}
\caption{Structural parameters of the two-component Sersic modeling for UGC~7346. Stellar masses are calculated using predictions by \cite{2015MNRAS.452.3209R} with the \textit{g} and \textit{r} bands.}
\label{tab:phot}
\end{center}
\end{table}

In the lower panels of Fig. \ref{fig:Panel_photo}, we show the two individual components of the double Sersic model. We refer to the components as 'main' and 'secondary', according to the established structural parameters, in which the main model dominates in luminosity and extension over the secondary model. Both models have coincident centrers. In Table~\ref{tab:phot} we show their structural properties. The colours of the main model are \textit{g}-\textit{r}~=~0.60 $\pm$~0.06~mag and \textit{r}-\textit{z}~=~0.27~$\pm$~0.12~mag, and for the secondary model, we have \textit{g}-\textit{r}~=~0.50~$\pm$~0.11~mag and \textit{r}-\textit{z}~=~0.64~$\pm$~0.22~mag. With this modelling, the structure of UGC~7346 is explained by a bluer component of smaller extent and luminosity that dominates the flux in the central region, while the dominant component is redder and of larger extent. The individual profiles of these main and secondary components of the double Sersic model (Fig.~\ref{fig:Panel_photo}) show how there is a transition in the colour profile from the approximate colour of the secondary model in the central region to the colour of the main component at higher radius. These circumstances confirm that the modeling with these two components is coherent and that the structure of UGC~7346 can be explained with two different stellar components with different and distinctive colours. 

While the models used for each component are synthetic Sersic models, there is an underlying substructure. This emerges in the images after subtracting the models from UGC~7346 and is also noticeable in the residual profile between UGC~7346 and the models (upper-right panel of Fig. \ref{fig:Panel_photo}). We find certain fluctuations up to a radius of 20 arcsec, which we associate with the streams or shells found in the central region of UGC~7346. We can also identify a bump in flux produced by a quasi-circular structure located at approximately 60 arcsec. We rule out that the residual structures located in the central region are a product of incorrect modeling, since they appear regardless of the fitted model and are even visible in the original image without model subtraction. On the other hand, we cannot rule out that the almost circular structure located at approximately 60 arcsec is some kind of artifact since it is located in the external regions, close to where the sky is oversubtracted in the data. Better and deeper imaging with no oversubtraction of the sky background will be needed to confirm or discard this outer stream.

The signs of interaction in the form of streams or shells, together with the existence of two stellar components with distinctive colours, indicate that UGC~7346 is most likely the result of a dwarf-dwarf merger. This statement is supported by previous work. \cite{2016MNRAS.456.1185B} showed from cosmological simulations that the product of dwarf-dwarf mergers is the presence of a bluer and more concentrated stellar component surrounded by a redder and older one in dwarf spheroidals. Additionally, \cite{2019ApJ...879...97C} showed that two blue-cored dwarfs in Virgo (as is the case of UGC~7346) are the product of dwarf-dwarf mergers by means of a kinematic analysis of their stellar component with spectroscopy. Therefore, the observational evidence we have clearly points to a 1:5 (see Table \ref{tab:phot}) dwarf-dwarf merger as responsible for the peculiar structure of UGC~7346.

\subsection{Globular cluster candidates}\label{sec:GCs}

\begin{figure*}
\centering
        \includegraphics[width=1.0\textwidth]{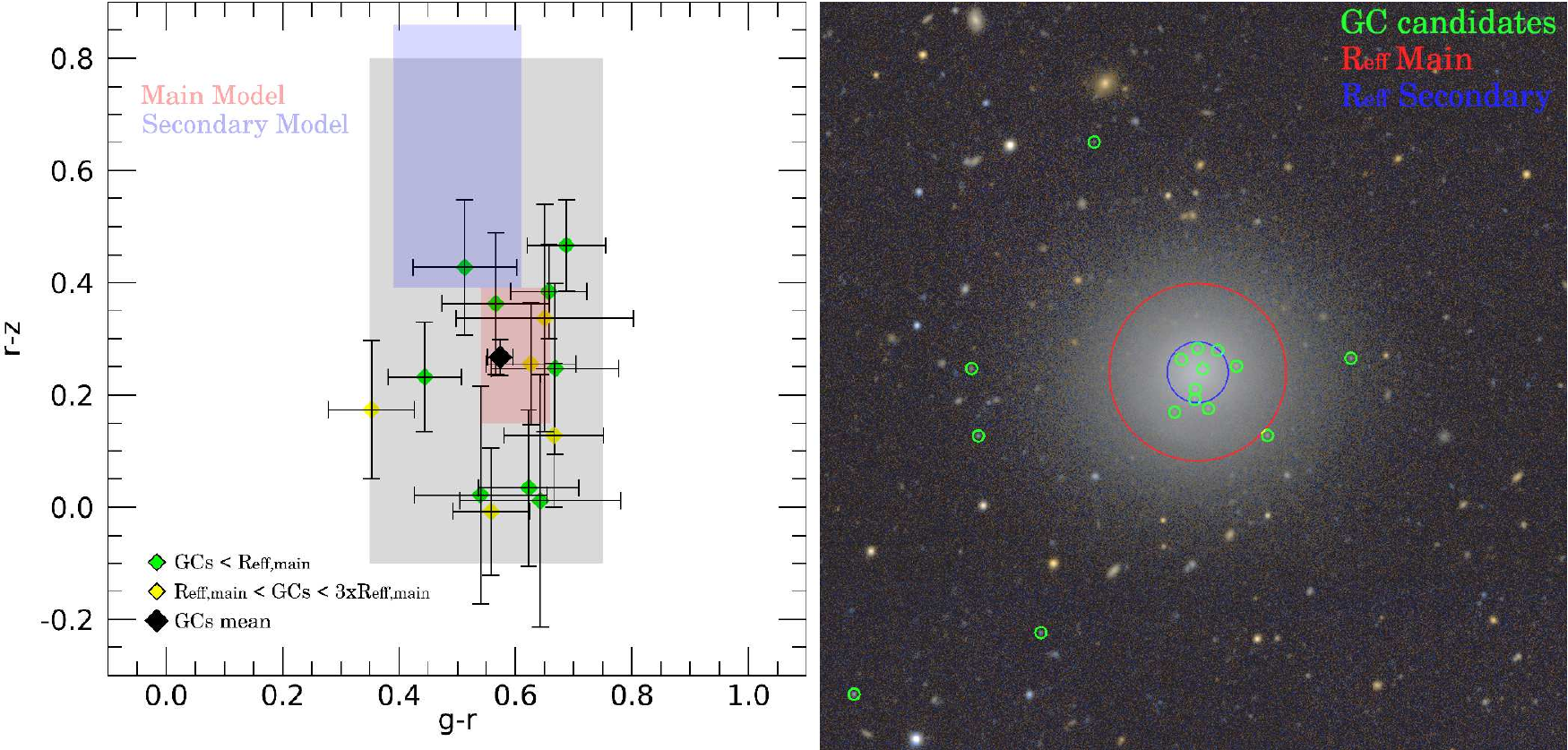}
    \caption{Analysis of GC candidates around UGC~7346. Left panel: \textit{g}-\textit{r} vs. \textit{r}-\textit{z} map showing the colour of detected GC candidates and the colours of the main and secondary Models (red and blue rectangles, respectively). The gray rectangle marks the colour space explored for GC candidate detection. GCs are differentiated by green colour (within one effective radius of the main model) and yellow colour (between one and three effective radii of the main model). Right panel: Composite colour image {(4.5 $\times$ 4.5 arcmin)} with \textit{g} and \textit{r} bands of UGC~7346. GC candidates detected by the selection criteria are plotted with green circles. The effective radii of the main and secondary models are indicated by the red and blue circles, respectively.
}
    \label{fig:GCs}
\end{figure*}

A significant concentration of point sources in the central region of UGC~7346 suggests the presence of a GC system. In order to make a catalogue of GC candidates, we carried out aperture photometry with the available \textit{g}, \textit{r,} and \textit{z} bands. This combination of bands is efficient for the detection of GC candidates \citep[see][who used similar data]{2021A&A...656A..44R}. In Appendix \ref{appendix:A}, we describe in the process for selecting point sources compatible with GC candidates, together with an analysis of the expected luminosity function of GCs for UGC~7346. An important point in this discussion is the fact that that we selected a range in the \textit{g}-\textit{r} vs. \textit{r}-\textit{z} colour space for the detection of GC candidates that is compatible with the colours of the two stellar components discussed in the previous section, namely, the potential GCs of these two stellar components.

In Fig. \ref{fig:GCs}, we show the results of this selection of GC candidates. In the left panel, we show the colour of the detected GC candidates around UGC~7346. The GC candidates are distributed in the region compatible with the colour of the main model. We indicate the average colour of all the {14} GC candidates located within three times the effective radius of the main model with a black symbol. We use a weighted mean taking into account photometric errors. The average colour of the GC candidates is only compatible with the main model, which confirms that the GCs belong to this stellar component. In the right panel, we show the distribution of the GC candidates in UGC~7346. We overplot the effective radius of both stellar components or models discussed above. The high concentration of GC candidates in the central region of UGC~7346 is discussed later in this work.

%We mark with a gray region the colour space used to select potential GC candidates. By red and blue regions we denote the average colour within errors of the Main and Secondary Models. We show with green symbols the positions of GC candidates located within the region of one effective radius of the Main Model. GC candidates located in the region between one and three times the effective radius of the Main Model are plotted with yellow symbols

\begin{figure}
\centering
        \includegraphics[width=1.0\columnwidth]{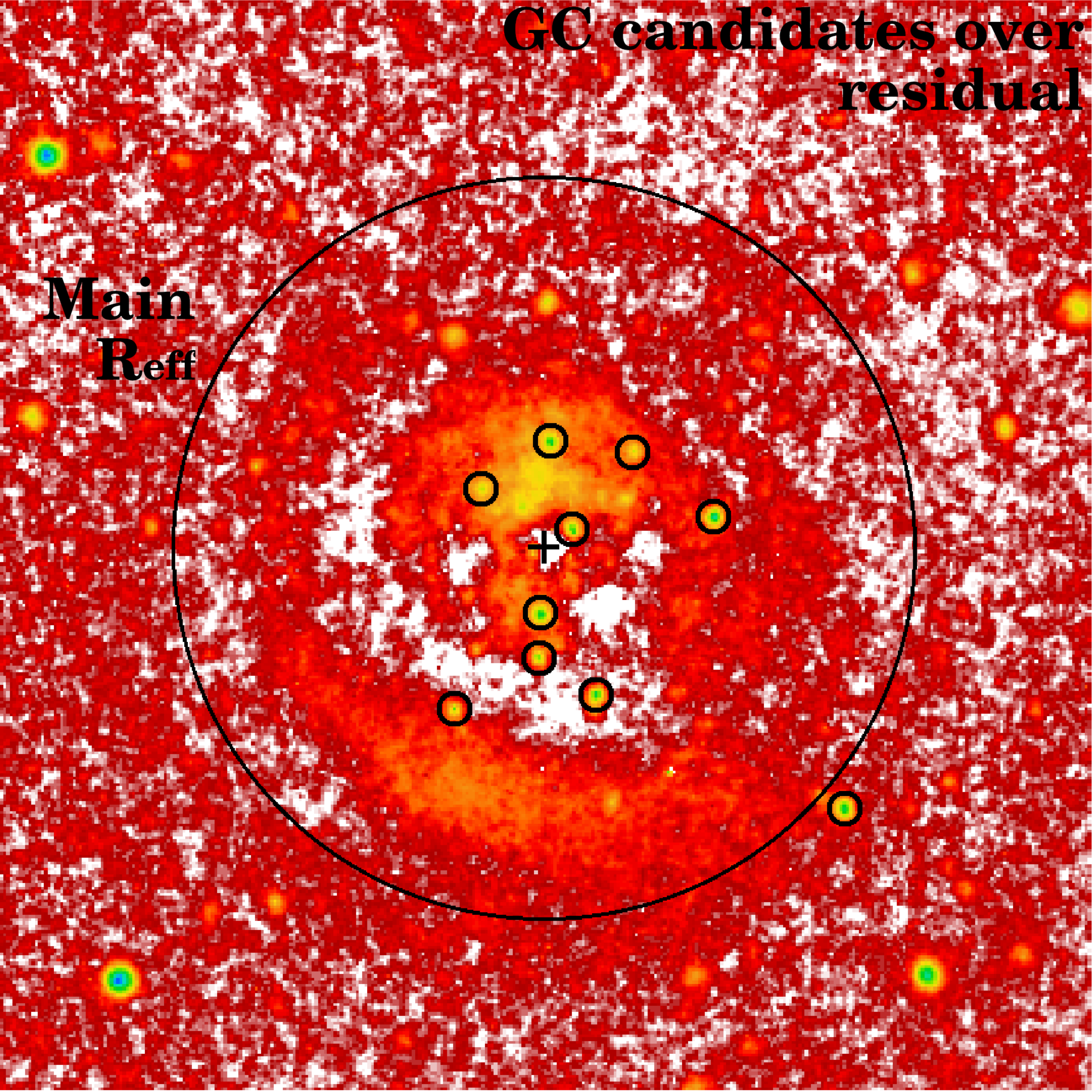}
    \caption{High-contrast view of the location of the GC candidates (small black circles) on the residual \textit{g}+\textit{r} image after subtracting the models discussed in Section \ref{sec:Models}. We mark the centre of UGC~7346 with a black cross. The stamp is 70 arcsec sideways over the central region of UGC~7346. The image has been smoothed with the Fully Adaptive Bayesian Algorithm for Data Analysis \citep[FABADA; ][]{2022arXiv220105145S} software, enhancing the diffuse emission while preserving the original resolution.}
    \label{fig:Res_GCs}
\end{figure}

In Fig. \ref{fig:Res_GCs}, we show an image similar to the right panel of Fig. \ref{fig:GCs}, but this time overplotting the GC candidates on the residual \textit{g}+\textit{r} image in high contrast after having subtracted the models and zooming in on the central region. We find that the position of the GCs coincides remarkably well with regions of flux overdensity in this residual image. Of the nine GC candidates located within the main model effective radius, eight of them are located where a residual positive flux is found after model subtraction.

\section{Discussion}

The analysis carried out in the previous section shows properties in UGC~7346 that are atypical in terms of two aspects. By decomposing the galaxy light of UGC~7346 we find that its morphology is explained by the superposition of two different stellar components with distinctly different colours (see Fig. \ref{fig:Panel_photo}). Additionally, by analyzing the residual images after the subtraction of the models we find what look like remnants of an interaction in the form of streams or shells. In this regard, we comment on the presence of substructures with spiral patterns in early-type dwarfs in the recent work by \cite{2021AJ....161..268M, 2022AJ....164...18M}. These structures are found underneath the main smooth component of dwarf galaxies in the Virgo and Fornax clusters;  \cite{2021ApJ...912..149S} have proposed that they are due to tidal forces causing an underlying disk component to emerge. 

We argue that the substructure found in UGC~7346 is qualitatively different from those structures. First, UGC~7346 has two clearly distinct stellar components and not only one dominant smooth component. In addition, the residual substructure in UGC~7346 is considerably more asymmetric than the early-type dwarf structures by \cite{2021AJ....161..268M, 2022AJ....164...18M}, which are fairly symmetric. We considered the possibility that the residuals could be a potential dust lane. The available optical SDSS spectra for UGC~7346\footnote{\url{http://skyserver.sdss.org/dr16/en/tools/explore/Summary.aspx?id=1237668624622944312}} show a passive population with no signs of star formation. We did not find  HI detection in UGC~7346 nor mid-IR counterparts to the features in the unWISE data \citep{2014AJ....147..108L}. These arguments together, including additional arguments discussed in Sec. \ref{sec:Models}, suggest that the overall structure of UGC~7346 is due to a dwarf-dwarf merger. This is in itself of high interest since few examples of early-type dwarf on-going mergers are found in the literature, with most dwarf-dwarf mergers showing high star formation activity \citep{2017NatAs...1E..25S, 2018ApJS..237...36P, 2020AJ....159..103K, 2020ApJ...900..152Z, 2021MNRAS.504.6179E}. 

Another remarkable aspect is the existence of a highly concentrated system of GCs with that colours that are broadly compatible with the main model component. Given the GC candidates found within three times the effective radius of the main model, we estimate a half-number radius (R$_{GC}$) for the GC system of approximately 13 arcsec. This implies R$_{GC}$ = 0.41 R$_{e}$, which is certainly a strong anomaly, since both the radial and azimuthal GC distributions follow those of the stars in dwarf galaxies, with typical values for R$_{GC}/$R$_{e}$ of the order of unity or above \citep[see][and references therein]{2022MNRAS.511.4633S}. We note that even assuming a single Sersic model to define the structure and effective radius of UGC~7346, the concentration of the GC candidates would be R$_{GC}$~=~0.50~R$_{e}$, so the very high concentration of the GC candidates is model-independent. 

We would like to point out that some of the specific properties of UGC~7346 have been previously observed  in other galaxies. For instance, the presence of blue cored dE in clusters is frequent and well known \citep{2017A&A...606A.135U, 2019A&A...625A..94H}. However, a complete analysis addressing both the presence of stellar subcomponents, the presence of substructure, and the distribution of globular clusters, does not exist in the literature. Considering the excellent data available in the nearby Virgo, Fornax and Coma clusters, this potential study could offer compatible cases to those we present in UGC~7346.

We suggest that the two coincident anomalies in UGC~7346, a dwarf-dwarf merger and the high concentration of its GCs, may be related to each other. A straightforward interpretation of the points discussed above would suggest that the dynamical friction produced by the ongoing dwarf-dwarf merger is collapsing the GCs towards the centre of UGC~7346. Indeed, simulations suggest that tidal forces in dwarf galaxies are sufficient to trigger a collapse of the GC system, {and \cite{2000ApJ...543..620O} were able to produce such a central concentration of GCs for some initial conditions}. This could be the case for UGC~7346. In principle, a merger event would provide a tidal field intense enough to create this collapse of the GCs by dynamical friction. However,  in the literature we did not
find a specific case of cosmological simulations of dwarf-dwarf mergers and their impact on the globular cluster system. 

To explore the hypothetical result of the merger of the GCs observed in the central region of UGC~7346, we calculated the total luminosity of this set of GCs, selecting those nine GC candidates that are inside the effective radius of the main model (see Fig. \ref{fig:Res_GCs}). The total integrated luminosity of this set of nine GC candidates is $L_{g}$~=~$-$10.2 mag. Considering the total mass of UGC~7346, $M_\star$~=~1.4~$\times$~10$^{9} M_\odot$, the luminosity of the hypothetical NSC formed by merging these GCs is in excellent agreement with the observed correlations between host galaxy mass and NSC luminosity \citep[see][]{2020A&ARv..28....4N}. There are, however, some caveats in this calculation. First, we do not know with certainty whether all or only some of the GC candidates in the selected region would eventually merge. Additionally, our photometric data are not complete to the entire GC luminosity function of UGC~7346. In Appendix \ref{appendix:A}, we show that the depth of our data allows us to map approximately half of the expected GCs in UGC~7346. Nevertheless, it is expected that these undetected GCs of low luminosity -- which could potentially merge as well -- would contribute little to the luminosity of the hypothetical NSC being formed. Therefore, we can conclude that the hypothetical NSC formed by the merging of the GC candidates found in the central region of UGC~7346 would have properties that are compatible with the NSCs and their hosts observed in the literature.

It is also interesting to explore the impact of dwarf-dwarf mergers as a viable channel for the formation of NSCs. Cosmological simulations show that dwarf-dwarf mergers are undoubtedly a common phenomenon \citep{2014ApJ...794..115D} and dwarfs typically undergo one major and one minor merger in their evolution \citep{2021MNRAS.500.4937M}. Assuming that dwarf merging is a viable channel to induce the collapse of GCs and the subsequent formation of an NSC, this would provide a natural explanation for the correlations between nucleation fraction and environmental density \citep[e.g., ][]{2019ApJ...878...18S, 2022A&A...664A.167S, 2022ApJ...927...44C}. In a straight forward argumentation, we can assume that a higher environmental density would produce a higher probability of interactions or merging between dwarf galaxies, thus explaining this correlation. The merging of two dwarfs could even explain the presence of young populations found in NSCs, since both stellar components of the minor component could potentially be younger, or the possible presence of gas could promote in situ star formation. While this scenario is somewhat speculative, we consider it of great interest to explore it in detail. Thus, future theoretical and observational works could confirm whether dwarf mergers do indeed serve as a viable channel for the formation of NSCs.

To conclude, we consider UGC~7346 to be an object of great interest for follow-up observations. First of all, our analysis of its GC system shows that we are limited by point source detections, so that deeper and higher resolution observations would show a larger number of GCs. Additionally, spectroscopic studies of both the GCs and the stellar components would offer the ability to provide dynamical information of the system and a better analysis of its stellar populations -- as keys to confirming the hypothesis that the fate of the GCs in UGC~7346 is to merge and ultimately produce a NSC.

\begin{acknowledgements}
{We thank the referee for a fair review of our work.} We acknowledge financial support from the State Research Agency (AEI-MCINN) of the Spanish Ministry of Science and Innovation under the grant "The structure and evolution of galaxies and their central regions" with reference PID2019-105602GB-I00/10.13039/501100011033, from the ACIISI, Consejer\'{i}a de Econom\'{i}a, Conocimiento y Empleo del Gobierno de Canarias and the European Regional Development Fund (ERDF) under grant with reference PROID2021010044, and from IAC project P/300724, financed by the Ministry of Science and Innovation, through the State Budget and by the Canary Islands Department of Economy, Knowledge and Employment, through the Regional Budget of the Autonomous Community. JR acknowledges funding from University of La Laguna through the Margarita Salas Program from the Spanish Ministry of Universities ref. UNI/551/2021-May 26, and under the EU Next Generation.
\end{acknowledgements}


\appendix

\section{Properties of GC candidates}\label{appendix:A}

\begin{figure*}
\centering
        \includegraphics[width=1.0\textwidth]{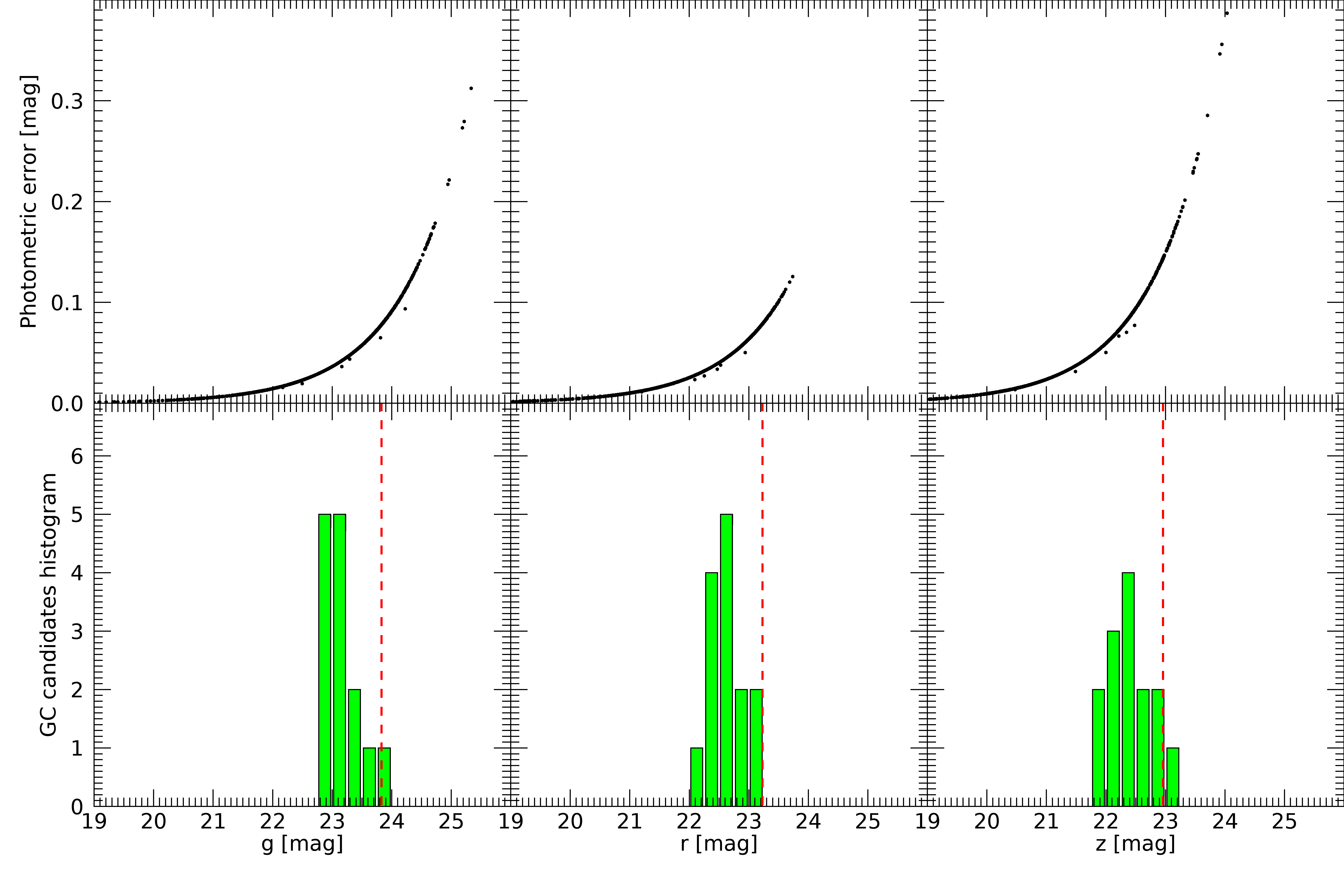}
    \caption{ Upper panels: Photometric errors in aperture photometry (see text) for sources in the field of UGC~7346 for the \textit{g}, \textit{r} and \textit{z} bands. Lower panels: Magnitude histograms for the GC candidates of UGC~7346. The expected peak of the GCLF is marked by a red dashed line.}
    \label{fig:apendix}
\end{figure*}

\begin{table*}
\begin{center}
\begin{tabular}{ccccc}
R.A. (J2000) & Dec. (J2000) & \textit{g} & \textit{r} & \textit{z} \\ 
{[deg]} & {[deg]} & {[mag]} & {[mag]} & {[mag]} \\
\hline
\multicolumn{5}{c}{GCs within 1 effective radius}\\
184.6764 & 17.7147 & 23.71 $\pm$ 0.07 & 23.07 $\pm$ 0.07 & 23.06 $\pm$ 0.16\\
184.6729 & 17.7150 & 23.19 $\pm$ 0.04 & 22.57 $\pm$ 0.04 & 22.53 $\pm$ 0.10\\
184.6743 & 17.7159 & 23.45 $\pm$ 0.06 & 22.91 $\pm$ 0.06 & 22.89 $\pm$ 0.13\\
184.6743 & 17.7169 & 22.76 $\pm$ 0.03 & 22.32 $\pm$ 0.03 & 22.08 $\pm$ 0.06\\
184.6757 & 17.7199 & 23.18 $\pm$ 0.04 & 22.67 $\pm$ 0.05 & 22.24 $\pm$ 0.07\\
184.6699 & 17.7193 & 22.90 $\pm$ 0.03 & 22.25 $\pm$ 0.03 & 21.86 $\pm$ 0.05\\
184.6735 & 17.7190 & 23.24 $\pm$ 0.05 & 22.68 $\pm$ 0.05 & 22.31 $\pm$ 0.08\\
184.6719 & 17.7208 & 23.48 $\pm$ 0.06 & 22.81 $\pm$ 0.05 & 22.56 $\pm$ 0.10\\
184.6740 & 17.7210 & 22.96 $\pm$ 0.03 & 22.27 $\pm$ 0.03 & 21.80 $\pm$ 0.05\\
\multicolumn{5}{c}{GCs within 1 and 3 effective radius}\\
184.6971 & 17.7123 & 22.88 $\pm$ 0.03 & 22.53 $\pm$ 0.04 & 22.36 $\pm$ 0.08\\
184.6667 & 17.7123 & 22.87 $\pm$ 0.03 & 22.31 $\pm$ 0.03 & 22.32 $\pm$ 0.08\\
184.6978 & 17.7190 & 23.21 $\pm$ 0.04 & 22.54 $\pm$ 0.04 & 22.41 $\pm$ 0.09\\
184.6578 & 17.7200 & 23.07 $\pm$ 0.04 & 22.45 $\pm$ 0.04 & 22.19 $\pm$ 0.07\\
184.6849 & 17.7417 & 23.83 $\pm$ 0.08 & 23.18 $\pm$ 0.07 & 22.84 $\pm$ 0.13\\
\hline
\newline
\end{tabular}
\caption{Coordinates and photometric magnitudes of the GC candidates in UGC~7346 located within three effective radii.}
\label{tab:GCs}
\end{center}
\end{table*}

In this section, we describe the process for selecting GC candidates and their properties. To screen GC candidates we use the model-subtracted images (discussed in Section \ref{sec:Models}) for cleaner point source photometry. We use \texttt{SExtractor} \citep{1996A&AS..117..393B} to perform photometry with an aperture of 2.1 arcsec, $\sim$2 times the FWHM of the data, which is 1.1 arcsec in the \textit{r} band. We used \texttt{SExtractor} in dual mode, with a detection threshold of 5$\sigma$, using the \textit{g}+\textit{r} image as the detection image to measure photometry in the individual \textit{g}, \textit{r} and \textit{z} bands. We filtered out point sources with the following criteria: 0.35 mag < \textit{g}-\textit{r} < 0.75 mag, $-$0.1 mag < \textit{r}-\textit{z} < 0.8 mag, ellipticity < 0.4, and FWHM < 1.35 arcsec.

In Fig. \ref{fig:apendix}, we plot the photometric errors for all detected sources in our data, as a function of their magnitude, in the upper panels for each of the \textit{g}, \textit{r,} and \textit{z} bands. In the lower panels, we plot the histogram of sources compatible with GC candidates within three times the effective radius of the main model, as discussed in Section \ref{sec:GCs}. Additionally, we indicate the expected position for the peak in magnitude of the GC luminosity function (GCLF) for an object located in the Virgo cluster, and thus at a distance of 16.5 Mpc (distance modulus of 31.09 mag). We used as our reference the value provided by \cite{2012Ap&SS.341..195R} in the \textit{V} band, adapted to the SDSS \textit{r} band by \cite{2019MNRAS.486..823R} for a diffuse dwarf of similar morphology and stellar mass to UGC~7346. To obtain the peak values of the GCLF in the \textit{g} and \textit{z} bands we used the colours of the main model, as discussed in Section \ref{sec:Models}.

Because the selection of GC candidates was performed using a residual $g+r$ sum image to calculate photometric magnitudes in \textit{g}, \textit{r,} and \textit{z} bands -- and additionally for a GC candidate to be selected it requires reliable photometry in all three bands -- a strict completeness analysis is very complex. However, the magnitude histograms of the selected GC candidates provide useful information. The vast majority of the GC candidates detected are in the bright half of the GCLF. We also find that the brightest GCs would be found at a distance of about 1 mag from the GCLF peak in the three bands, \textit{g}, \textit{r,} and \textit{z}. This is as expected since, for the stellar mass of UGC~7346, one sigma of the GCLF will be on the order of (or below) 1 mag \citep{2007ApJS..171..101J, 2010ApJ...717..603V}. This provides additional evidence that we are indeed detecting GCs in UGC~7346.

\end{document}